\documentclass[a4paper,11pt]{article}
\usepackage{pos}

\title{Measurement of the top quark pole mass using \ttbarjet\ events in the dilepton final state at $\sqrt{s}=$ 13 TeV with the CMS experiment}
\ShortTitle{Measurement of the top quark pole mass using \ttbarjet\ events in CMS}
\author*[1]{Sebastian Wuchterl}

\affiliation{Deutsches Elektronen-Synchrotron (DESY),\\
  Notkestrasse 85, Hamburg, Germany}
\note{For the CMS Collaboration.}
\emailAdd{sebastian.wuchterl@cern.ch}

\abstract{
In this contribution, a measurement of the top quark pole mass \mtpole\ is presented. Events where a top quark-antiquark pair (\ttbar) is produced in association with at least one additional energetic jet (\ttbarjet) are analyzed. The data set recorded by the CMS experiment at a center-of-mass energy of 13 TeV, corresponding to a total integrated luminosity of 36.3\,\fbinv, is used. Reconstructed events are required to contain two opposite-sign leptons in the final state.
Machine learning methods are employed for the reconstruction of the main observable event classification.
The normalized \ttbarjet\ production cross section as a function of the inverse of the invariant mass of the \ttbarjet\ system is measured at the parton level using a likelihood unfolding method.
The value for \mtpole\ is extracted by comparing the measurement to next-to-leading order theoretical predictions using two parton distribution functions (PDFs). For a reference set of the ABMP16NLO PDF, the resulting value is $\mtpole = 172.94\pm1.37\,\text{GeV}$.
}

\FullConference{%
  The Tenth Annual Conference on Large Hadron Collider Physics - LHCP2022\\
  16-20 May 2022\\
  online
}

\newcommand{\ttbar}{\ensuremath{\mathrm{t\overline{t}}}}
\newcommand{\ttbarjet}{\ensuremath{\mathrm{t\overline{t}}\text{+jet}}}
\newcommand{\ttbarnojet}{\ensuremath{\mathrm{t\overline{t}}\text{+0jet}}}
\newcommand{\fbinv}{\ensuremath{\mathrm{fb^{-1}}}}

\newcommand{\mt}{\ensuremath{\mathrm{m_{t}}}}
\newcommand{\mtpole}{\ensuremath{\mathrm{m_{t}^{pole}}}}
\newcommand{\mtmc}{\ensuremath{m_\mathrm{{t}}^\mathrm{{MC}}}}
\newcommand{\pt}{\ensuremath{p_\mathrm{{T}}}}

\begin{document}
\renewcommand{\logo}{\relax}
\maketitle

\section{Introduction}
The top quark mass \mt\ is a free parameter of the standard model (SM) and its large mass value indicates that it plays a particular role in the SM. It is also of high relevance for global electroweak fits. Because \mt\ has to be determined experimentally, it motivates complementary methods for its measurement.
Direct measurements of \mt, which rely on variables sensitive to the reconstructed energy of the top quark and are performed using template fits using multi-purpose Monte-Carlo (MC) simulation, reach a precision on the order of 0.5 GeV. The downside is the dependence on the modeling of non-perturbative effects by heuristic models tuned to experimental data. This leads to an interpretation problem when compared to well-defined masses, such as the top quark pole mass \mtpole, and is commonly covered by an uncertainty on the order of 0.5--1 GeV.

In the presented measurement, the value of \mtpole\ is extracted using the normalized differential cross section of top quark-antiquark pair (\ttbar) production in association with at least one additional jet (\ttbarjet) by the CMS Collaboration~\cite{CMS-PAS-TOP-21-008}. The cross section at parton level is measured as a function of the $\rho$ observable, defined as $340\,\text{GeV}/m_{\ttbarjet}$, where $m_{\ttbarjet}$ is the invariant mass of the \ttbarjet\ system. The additional jet needs to have a transverse momentum of \pt\ > 30 GeV and an absolute pseudorapidity of $|\eta|<2.4$.
High mass sensitivity is expected for $\rho>0.65$, close to the production threshold.
The definition and analysis strategy follows the approach suggested in Refs.~\cite{bib:runningMass3,bib:ttjet1,bib:ttjet2} and is the first time performed at a center-of-mass energy of 13 TeV. Similar measurements~\cite{bib:poleMass4,bib:ttjetAtlas8} were performed previously by the ATLAS Collaboration, yielding a value of $\mtpole = 171.1\,^{+1.2}_{-1.0}\,\text{GeV}$.

\section{Analysis setup and strategy}
The data sample analyzed was recorded by the CMS experiment and corresponds to an integrated luminosity of 36.3\,\fbinv. Events are selected based on a combination of single lepton and dilepton triggers, where events are retained in the offline selection if they contain at least one opposite-sign lepton pair (ee, e$\mathrm{\mu}$, $\mathrm{\mu\mu}$) with \pt\ > 25 (20) GeV for the leading (subleading) lepton. Jets are selected with \pt\ > 30 GeV, and both leptons and jets are required to be reconstructed within $|\eta|<2.4$. Jets originating from the hadronization of b quarks (b jets) are identified using a dedicated tagging algorithm.
Additional selection requirements are imposed to suppress background contributions from Z+jets production.
Finally, events are selected if they contain at least one reconstructed b jet.

\begin{figure}[htbp]
\centering
\includegraphics[width=0.45\textwidth]{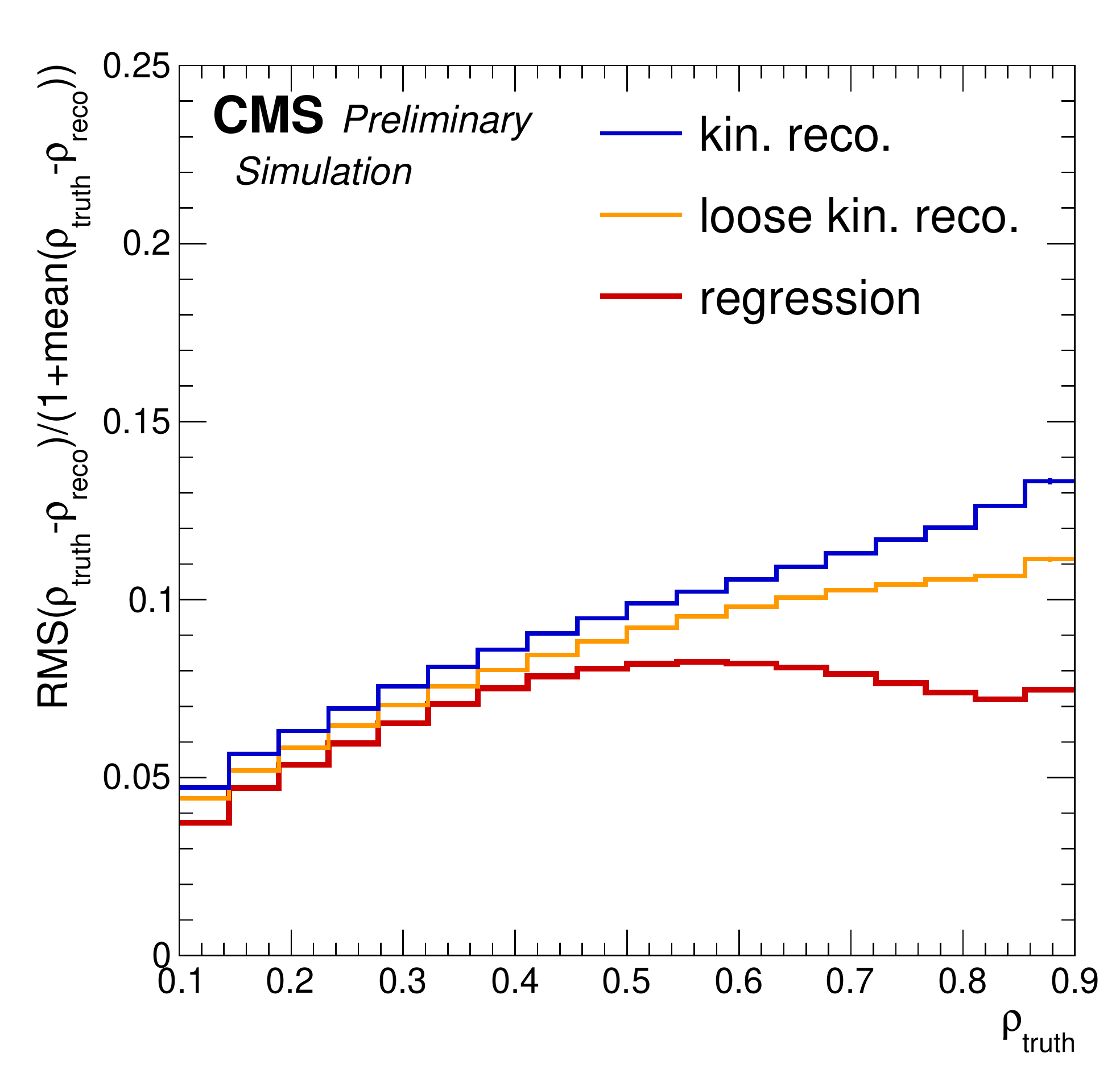}
\includegraphics[width=0.45\textwidth]{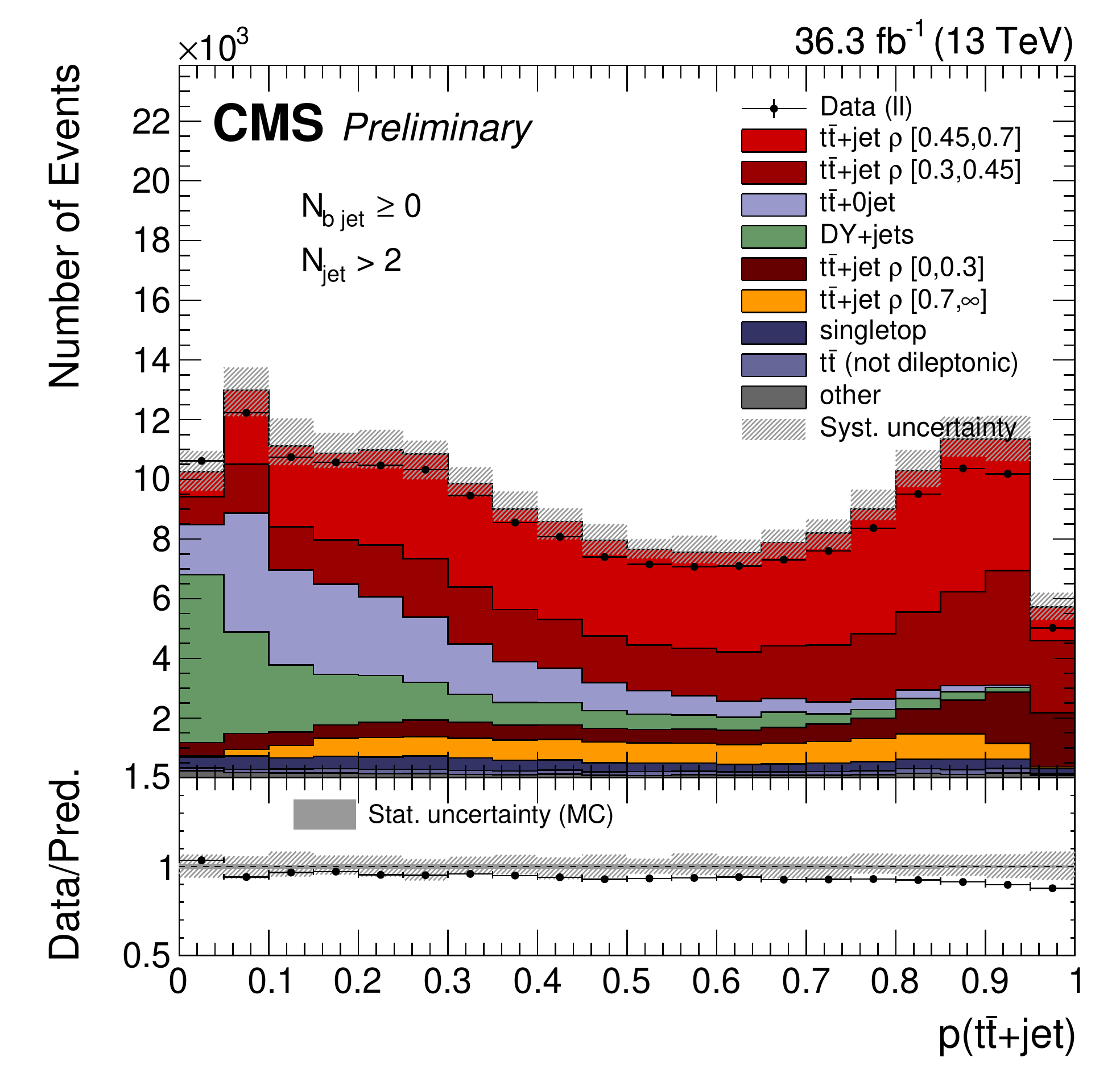}
\caption{The $\rho$ resolution as a function of the truth value for the NN regression and two analytical methods (left)~\cite{CMS-PAS-TOP-21-008}. The observed and MC predicted signal and background yields as a function of the signal output node of the classifier NN~\cite{CMS-PAS-TOP-21-008}.}
\label{fig:NN}
\end{figure}

Since a significant amount of missing transverse momentum arises for dileptonic \ttbar\ events due to the escaping neutrinos, this leads to challenges in the kinematic reconstruction of the $\rho$ variable. Thus, a dedicated neural network (NN) based regression method is developed to improve the efficiency and resolution with respect to classical analytical approaches. The resolution for the reconstructed value of $\rho$ is shown in Fig.~\ref{fig:NN} left, where the NN regression is compared to two approaches used in previous CMS measurements. A significant improvement is visible, especially for large values of $\rho$.

The dominant background contribution arises from \ttbar\ production without an additional jet (\ttbarnojet). The additional jet can originate from pileup effects or migrations due to the limited jet energy resolution. This leads to further challenges as the kinematic event properties of \ttbarnojet\ and \ttbarjet\ are similar, and the background is mainly dominant for large values of $\rho$. Hence, an NN-based classifier is developed, and events are classified to originate from either \ttbarjet, \ttbarnojet, or Z+jets production. The multiclass classification network is calibrated such that the performance is independent of the value of $\rho$. The response for the \ttbarjet\ output node of the NN is shown in Fig.~\ref{fig:NN} right, where the data is compared to the MC predictions.

\begin{figure}[htbp]
\centering
\includegraphics[width=0.9\textwidth]{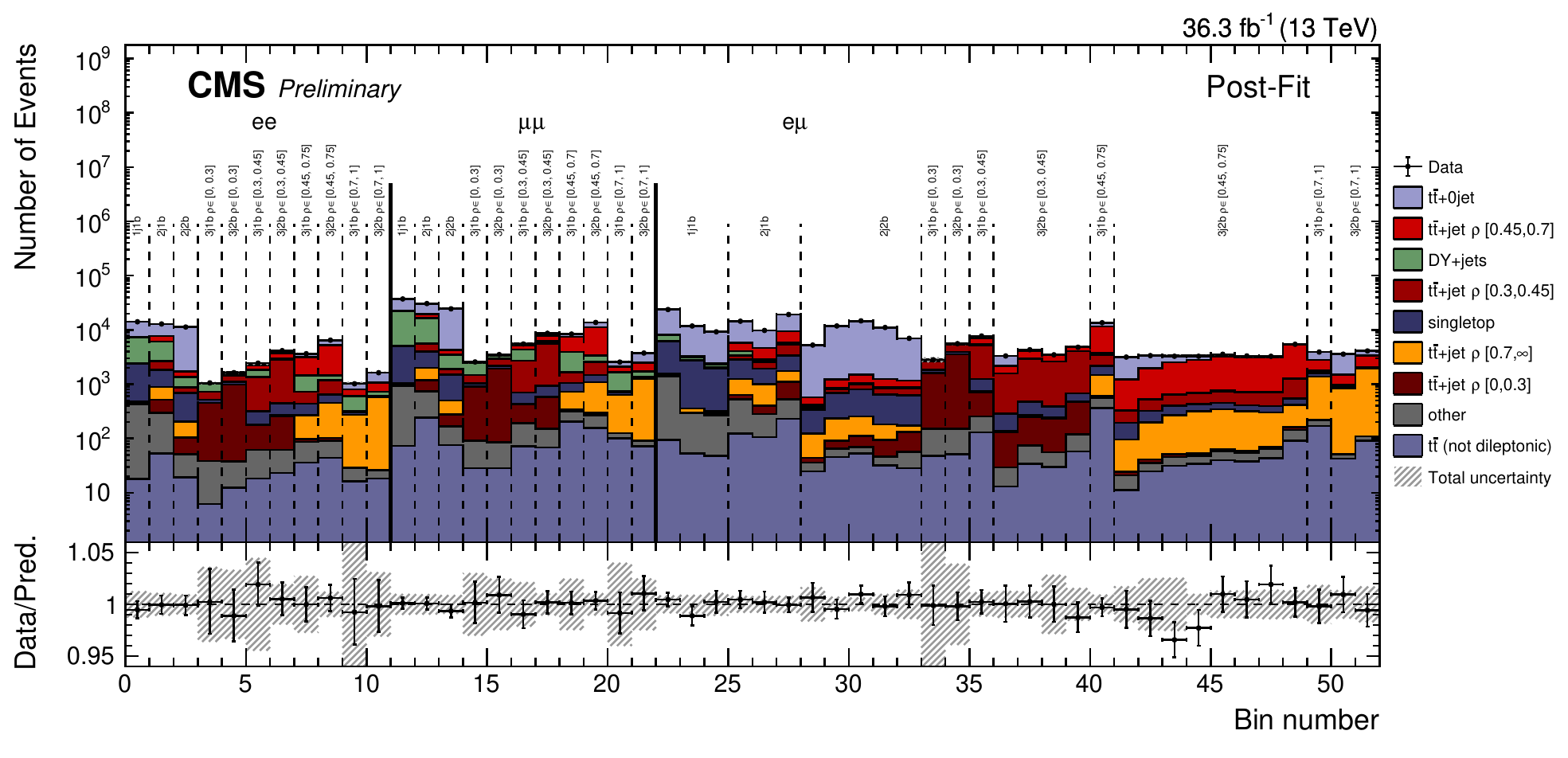}
\caption{The measured data and MC predicted signal and background yields after the likelihood fit. The gray band indicates the total uncertainty after the fit to the data~\cite{CMS-PAS-TOP-21-008}.}
\label{fig:postfit}
\end{figure}

The parton level \ttbarjet\ cross section as a function of $\rho$ is directly obtained from the result of a multidimensional likelihood fit to final-state observables. Events are separated based on the reconstructed value of $\rho$, the jet and b jet multiplicities, and the lepton channel.
The binning of $\rho$ at the detector and parton level follows studies on the diagonality of the response matrix and bin-to-bin migrations. It is determined to be: 0--0.3, 0.3--0.45, 0.45--0.7,and 0.7--1. The likelihood function is constructed based on a Poisson description of the observed event yields, and systematic uncertainties are incorporated as nuisance parameters. Additionally, a free parameter is introduced for the top quark mass assumed in the MC simulation, \mtmc, to reduce its dependence as it enters the modeling of the acceptance and the extrapolation to the full phase space.
The observables used in the fit are the relative response of the NN classifier with respect to the \ttbarnojet\ background, the \pt\ of the lowest energetic jets, the invariant mass of the lepton and b jet pair, and the absolute event yield. This is to maximize the measurement precision while retaining sensitivity to dominant systematic uncertainties and \mtmc. The distributions of the signal and background MC predictions after the fit to the data are shown in Fig.~\ref{fig:postfit}. A good agreement is observed for all categories and bins.

\section{Results and top quark mass extraction}
\begin{figure}[htbp]
\centering
\includegraphics[width=0.45\textwidth]{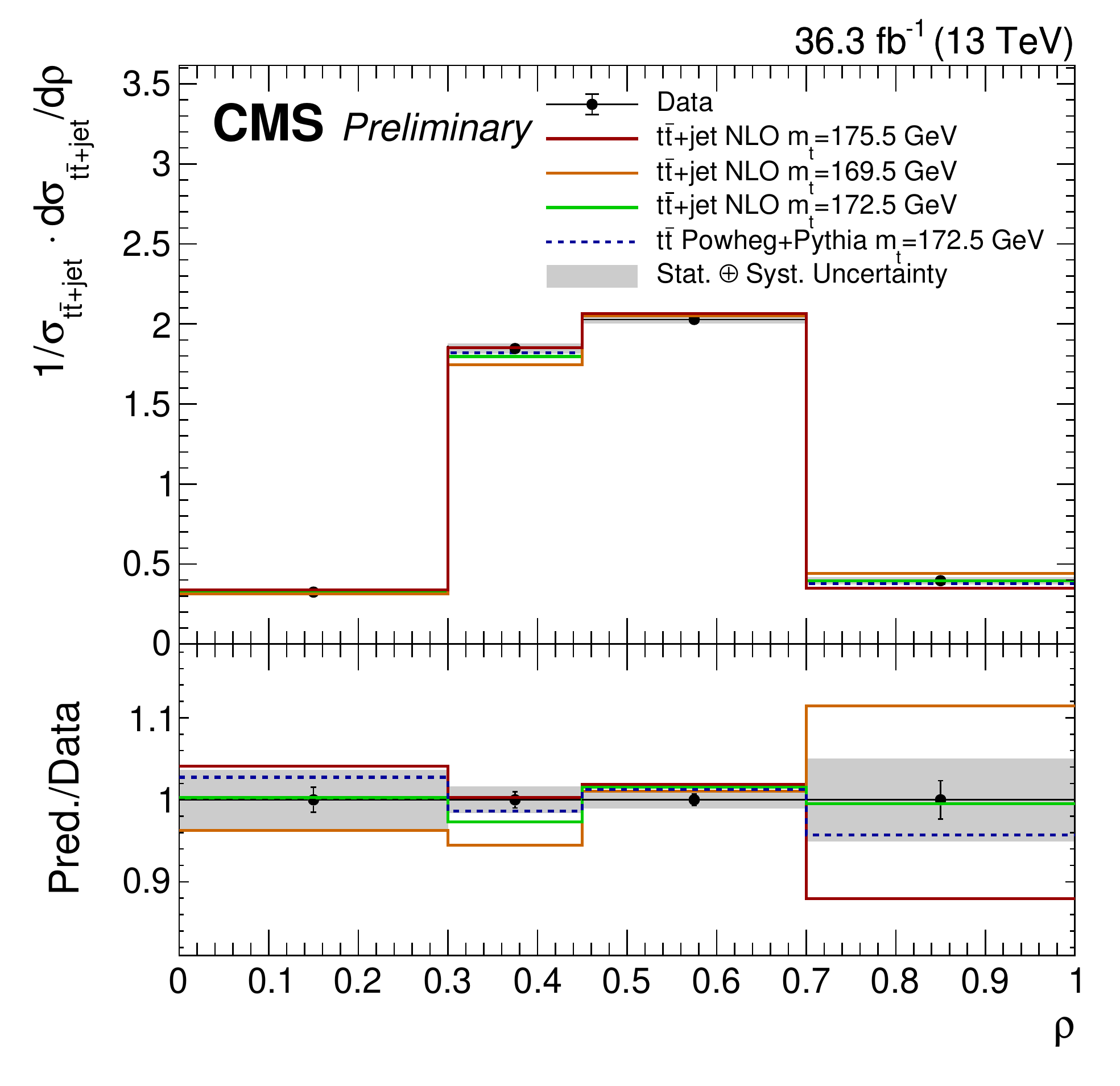}
\includegraphics[width=0.45\textwidth]{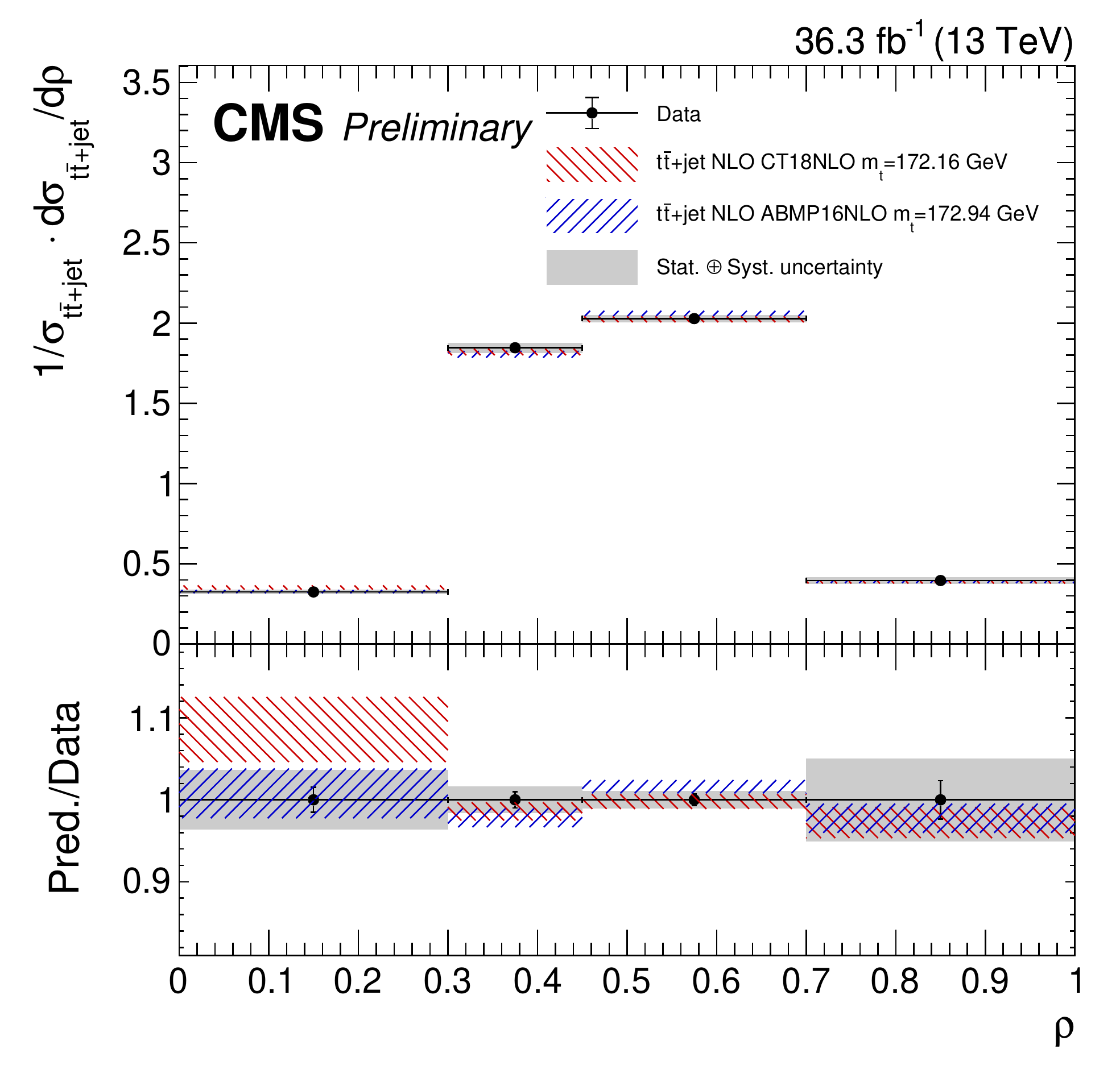}
\caption{The normalized \ttbarjet\ differential cross section as a function of $\rho$, compared to NLO theoretical predictions assuming different top quark mass values (left)~\cite{CMS-PAS-TOP-21-008}. The same measured cross section is shown but now compared to the theoretical predictions using the best-fit values for \mtpole\ (right)~\cite{CMS-PAS-TOP-21-008}.}
\label{fig:results}
\end{figure}

The measured normalized parton level \ttbarjet\ cross section is shown in Fig.~\ref{fig:results} left, comparing it to next-to-leading-order (NLO) theoretical predictions~\cite{bib:ttjPheno}. Different values for \mtpole\ are used for the NLO prediction to indicate the mass sensitivity. Here, the ABMP16NLO~\cite{bib:ABMP16} parton distribution function (PDF) is considered.
The value for \mtpole\ is extracted from a $\chi^2$ fit of the normalized differential cross section to the NLO predictions. In the covariance matrix used in the fit, all bin-to-bin correlations, as well as PDF uncertainties and uncertainties arising from the extrapolation to the full phase space are fully taken into account. The uncertainty due to the choice of the matrix-element scales is treated externally.
A value for \mtpole\ of $172.94\pm1.27\,\text{(fit)}\,^{+0.51}_{-0.43}\,\text{(scale)}\,\text{GeV}$ is determined. When using the CT18NLO~\cite{bib:CT18} PDF set, the value is $172.16\pm1.35\,\text{(fit)}\,^{+0.50}_{-0.40}\,\text{(scale)}\,\text{GeV}$. The measured cross section compared to the best-fit predictions is shown in Fig.~\ref{fig:results} right.

\bibliographystyle{JHEP}
\bibliography{main}

\providecommand{\href}[2]{#2}\begingroup\raggedright\begin{thebibliography}{1}

\bibitem{CMS-PAS-TOP-21-008}
{CMS Collaboration}, \emph{{Measurement of the top quark pole mass using
  $\text{t}\overline{\text{t}}\text{+jet}$ events in the dilepton final state
  at $\sqrt{s}=$ 13 TeV}},  CMS Physics Analysis Summary
  \href{https://cds.cern.ch/record/2804936}{CMS-PAS-TOP-21-008}, CERN, Geneva
  (2022).

\bibitem{bib:runningMass3}
J.~Fuster, A.~Irles, D.~Melini, P.~Uwer and M.~Vos, \emph{Extracting the
  top-quark running mass using $\ttbar+1$-jet events produced at the {Large
  Hadron Collider}},
  \href{https://doi.org/10.1140/epjc/s10052-017-5354-z}{\emph{Eur. Phys. J. C}
  {\bfseries 77} (2017) 794}
  [\href{https://arxiv.org/abs/1704.00540}{{\ttfamily 1704.00540}}].

\bibitem{bib:ttjet1}
S.~Alioli, P.~Fernandez, J.~Fuster, A.~Irles, S.~Moch, P.~Uwer et~al., \emph{A
  new observable to measure the top-quark mass at hadron colliders},
  \href{https://doi.org/10.1140/epjc/s10052-013-2438-2}{\emph{Eur. Phys. J. C}
  {\bfseries 73} (2013) 2438}
  [\href{https://arxiv.org/abs/1303.6415}{{\ttfamily 1303.6415}}].

\bibitem{bib:ttjet2}
G.~Bevilacqua, H.B.~Hartanto, M.~Kraus, M.~Schulze and M.~Worek, \emph{Top
  quark mass studies with $\ttbar\text{j}$ at the {LHC}},
  \href{https://doi.org/10.1007/JHEP03(2018)169}{\emph{JHEP} {\bfseries 03}
  (2018) 169} [\href{https://arxiv.org/abs/1710.07515}{{\ttfamily
  1710.07515}}].

\bibitem{bib:poleMass4}
{ATLAS Collaboration}, \emph{Determination of the top-quark pole mass using
  $\ttbar+1$-jet events collected with the {ATLAS} experiment in {7 TeV} pp
  collisions}, \href{https://doi.org/10.1007/JHEP10(2015)121}{\emph{JHEP}
  {\bfseries 10} (2015) 121}
  [\href{https://arxiv.org/abs/1507.01769}{{\ttfamily 1507.01769}}].

\bibitem{bib:ttjetAtlas8}
{ATLAS Collaboration}, \emph{Measurement of the top-quark mass in
  $\ttbar+1$-jet events collected with the {ATLAS} detector in pp collisions at
  $\sqrt{s} = {8\text{TeV}}$},
  \href{https://doi.org/10.1007/JHEP11(2019)150}{\emph{JHEP} {\bfseries 11}
  (2019) 150} [\href{https://arxiv.org/abs/1905.02302}{{\ttfamily
  1905.02302}}].

\bibitem{bib:ttjPheno}
S.~Alioli, J.~Fuster, M.V.~Garzelli, A.~Gavardi, A.~Irles, D.~Melini et~al.,
  \emph{Phenomenology of $\ttbar\mathrm{j}+\mathrm{X}$ production at the
  {LHC}}, \href{https://doi.org/10.1007/JHEP05(2022)146}{\emph{JHEP} {\bfseries
  05} (2022) 146} [\href{https://arxiv.org/abs/2202.07975}{{\ttfamily
  2202.07975}}].

\bibitem{bib:ABMP16}
S.~Alekhin, J.~Bl{\"u}mlein and S.~Moch, \emph{{NLO PDFs} from the {ABMP16
  fit}}, \href{https://doi.org/10.1140/epjc/s10052-018-5947-1}{\emph{Eur. Phys.
  J. C} {\bfseries 78} (2018) 477}
  [\href{https://arxiv.org/abs/1803.07537}{{\ttfamily 1803.07537}}].

\bibitem{bib:CT18}
T.-J.~Hou, K.~Xie, J.~Gao, S.~Dulat, M.~Guzzi, T.J.~Hobbs et~al., ``Progress in
  the {CTEQ-TEA NNLO global QCD analysis}.'' 2019.

\end{thebibliography}\endgroup



\end{document}